\documentclass
[aps,prl,amsfonts,amssymb,twocolumn,amsmath,preprintnumbers,nofootinbib,floatfix,
showpacs,superscriptaddress]{revtex4-1}%
\usepackage[dvips]{graphics}
\usepackage{graphicx}
\usepackage{dcolumn}
\usepackage{bm}
\usepackage{amsmath}
\usepackage{amsfonts}
\usepackage{amssymb}
\usepackage{xcolor}
\usepackage{subfigure}
\usepackage{hyperref,hypcap}
\usepackage{braket}
\usepackage{commath}%
\providecommand{\U}[1]{\protect\rule{.1in}{.1in}}

\begin{document}

\title{Temperature dependence of side-jump spin Hall conductivity}
\author{Cong Xiao}
\email{congxiao@utexas.edu}
\affiliation{Department of Physics, The University of Texas at Austin, Austin, Texas 78712, USA}

\author{Yi Liu}
\affiliation{The Center for Advanced Quantum Studies and Department of Physics, Beijing Normal University,
100875 Beijing, China}

\author{Zhe Yuan}
\email{zyuan@bnu.edu.cn}
\affiliation{The Center for Advanced Quantum Studies and Department of Physics, Beijing Normal University,
100875 Beijing, China}

\author{Shengyuan A. Yang}
\affiliation{Research Laboratory for Quantum Materials, Singapore University of Technology and Design, Singapore 487372, Singapore}

\author{Qian Niu}
\affiliation{Department of Physics, The University of Texas at Austin, Austin, Texas 78712, USA}

\begin{abstract}
In the conventional paradigm of the spin Hall effect, the side-jump
conductivity due to electron-phonon scattering is regarded to be temperature
independent. To the contrary, we draw the distinction that, while this
side-jump conductivity is temperature independent in the classical
equipartition regime where the longitudinal resistivity is linear in
temperature, it is temperature dependent below the equipartition regime. The
mechanism resulting in this temperature dependence differs from the familiar
one of the longitudinal resistivity. In the concrete example of Pt, we show
that the change of the spin Hall conductivity with temperature can be as high
as 50\%. Experimentally accessible high-purity Pt is proposed to be suitable
for observing this prominent variation below 80 K.
\end{abstract}
\maketitle

The spin Hall effect refers to a transverse spin current in response to an
external electric field \cite{Sinova2015}. In strongly spin-orbit-coupled
electronic systems such as 4$d$ and 5$d$ transition metals
\cite{Hoffmann2013,Xia2016,Tanaka2008,Sagasta2016,Jin2019,Seki2008,Hoffmann2010,Liu2012,Stamm2017}
and Weyl semimetals \cite{Yan2016}, the spin Hall conductivity due solely to
the geometry of Bloch bands, the so-called spin Berry curvature, has attracted
much attention. Besides, there is a scattering induced mechanism called
side-jump, whose contribution to the spin Hall conductivity turns out to be of
zeroth order of scattering time and independent of the density of a given type
of impurities \cite{Sinova2015}. Furthermore, the side-jump spin Hall
conductivity arising from the electron-phonon scattering is conventionally
regarded to be temperature ($T$) independent although the phonon density
varies with $T$ \cite{Berger1970,Bruno2001}.

In this work we draw the distinction that, while the electron-phonon
scattering induced side-jump spin Hall conductivity is $T$-independent in the
classical equipartition regime where the longitudinal resistivity $\rho$ is
linear in $T$, it is $T$-dependent at temperatures below the equipartition
regime. This character distinguishes side-jump from the geometric
contribution, and provides a new mechanism for $T$-dependent spin Hall
conductivities in high-purity experimental samples. An intuitive picture is
proposed for the $T$-dependence of the side-jump conductivity, which differs
from $\rho$ that is always $T$-dependent. Moreover, our first-principles
calculation demonstrates a prominent $T$-variation of the spin Hall
conductivity in experimentally accessible high-purity Pt below 80 K.

We consider strongly spin-orbit coupled multiband systems. The Fermi energy
and the interband-splitting around the Fermi level are assumed to be much
larger than the room temperature, thus the thermal smearing of Fermi surface
is negligible. Aiming to provide semi-quantitative and intuitive
understanding, the electron-phonon scattering is approximated by a
single-electron elastic process, which can be called the \textquotedblleft
quasi-static approximation\textquotedblright. In calculating the resistivity
resulting from phonon scattering, this approximation produces not only the
correct low-$T$ power law ($\rho\sim T^{5}$ for three-dimensional isotropic
single-Fermi-surface systems) \cite{Ziman1972} but also the values that are
quantitatively comparable with experimental data \cite{Liu2015}. When applied
to the side-jump transport, the high-$T$ and low-$T$ asymptotic behaviors are
grasped in this approximation. Quantitative deviations appearing in the
intermediate temperature regime are not essential for the present purpose.

The side-jump was originally proposed as the side-way shift in opposite
transverse directions for the carriers with different spins, when they are
scattered by spin-orbit active impurities \cite{Berger1970,Lyo1972}. This
picture works well in systems with weak spin-orbit coupling
\cite{Engel2005,Sarma2006,Vignale2006}, where the spin-orbit-induced band
splitting is smeared by disorder broadening \cite{Sinova2015}. Whereas in
strongly spin-orbit-coupled Bloch bands of current interest, the side-jump
contribution arises microscopically from the scattering-induced
band-off-diagonal elements of the out-of-equilibrium density matrix
\cite{KL1957,Sinitsyn2006,Sinitsyn2008,Xiao2019NLHE}. This corresponds to in the
Boltzmann transport formalism the dressing of Bloch states by interband
virtual scattering processes involving off-shell states away from the Fermi
surface \cite{Xiao2017SOT-SBE}.

\begin{figure}[tbh]
\includegraphics[width=1.0\columnwidth]{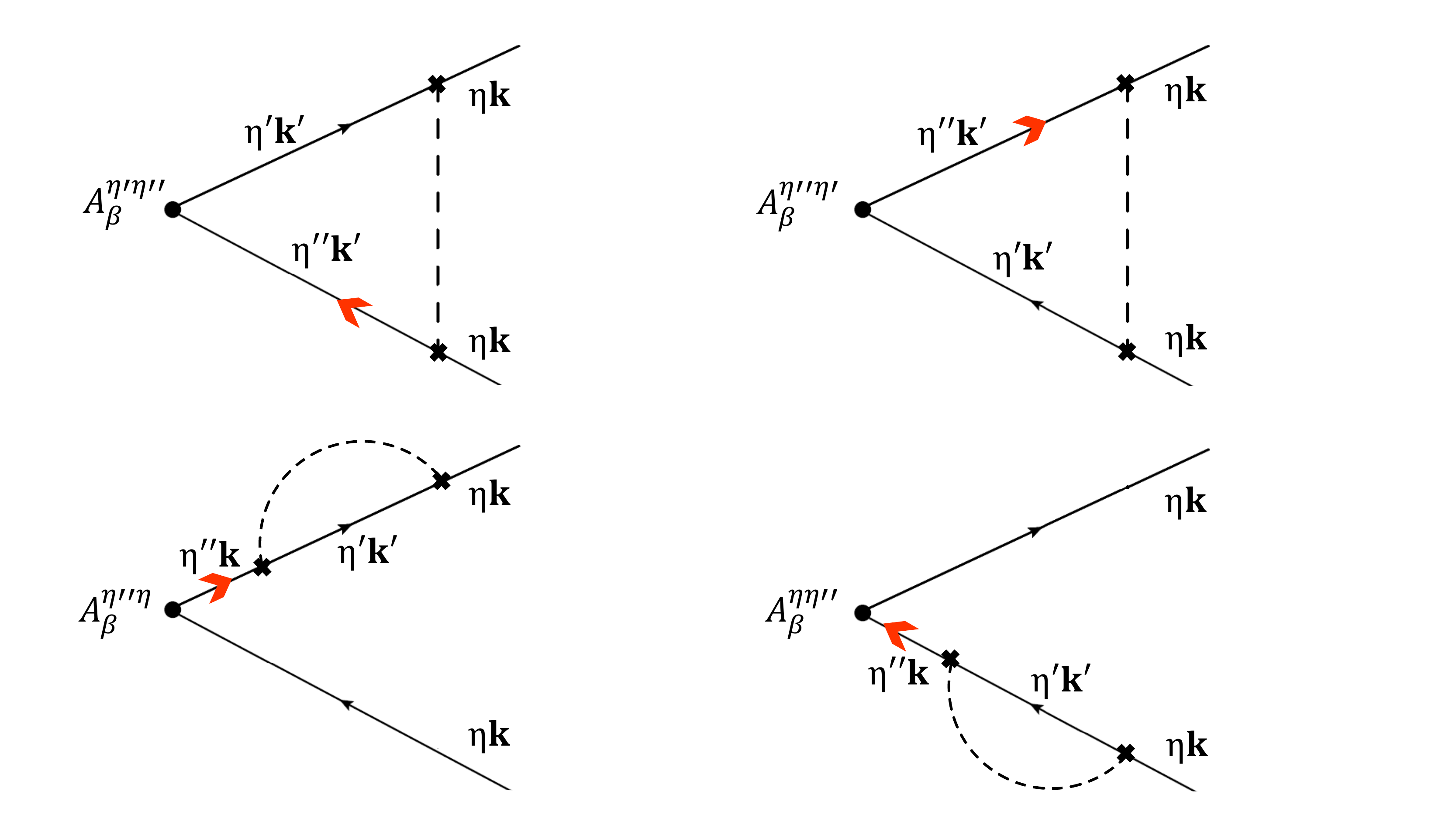}\caption{{}Graphical
representation of Eq. (\ref{sj-velocity}), where the off-shell states
away from the Fermi surface are marked by red arrows.
$W_{\mathbf{kk}^{\prime}}$ is represented by the
disorder line connected with two interaction vertexes.}%
\label{fig:sj-vertex}%
\end{figure}

\emph{{\color{blue} Transport formalism involving off-shell states.}}---In
weakly disordered crystals perturbed by an weak external electric field
$\mathbf{E}$, the expectation value of an observable $\mathbf{A}$ (assumed to
be a vector without loss of generality) reads%
\begin{equation}
\left\langle \mathbf{A}\right\rangle =\sum_{l}\mathbf{A}_{l}f_{l}%
\end{equation}
in the Boltzmann transport formalism, where $f_{l}$ is the occupation function
of the carrier state marked by $l=\left(  \eta,\mathbf{k}\right)  $ with
$\eta$ the band index and $\mathbf{k}$ the crystal momentum, $\mathbf{A}_{l}$
is the quantum mechanical average on state $l$. The carrier state is the Bloch
state dressed by interband virtual processes induced by both the electric
field and scattering \cite{Xiao2017SOT-SBE}. In the linear response and weak
scattering regime, these two dressing effects are independent
\cite{Xiao2017SOT-SBE}:%
\begin{equation}
\mathbf{A}_{l}=\mathbf{A}_{l}^{0}+\mathbf{A}_{l}^{\text{bc}}+\mathbf{A}%
_{l}^{\text{sj}}, \label{semi-A-inter}%
\end{equation}
where
\begin{equation}
\left(  \mathbf{A}_{l}^{\text{bc}}\right)  _{\beta}=\frac{e}{\hbar
}\mathbf{E}_{\alpha}\Omega_{\alpha\beta}^{A}\left(  \eta\mathbf{k}\right)
\end{equation}
arises from the electric-field induced dressing, with%
\begin{equation}
\Omega_{\alpha\beta}^{A}\left(  \eta\mathbf{k}\right)  =-2\hbar^{2}%
\operatorname{Im}\sum_{\eta^{\prime\prime}\neq\eta}\frac{v_{\alpha}^{\eta
\eta^{\prime\prime}}\left(  \mathbf{k}\right)  A_{\beta}^{\eta^{\prime\prime
}\eta}\left(  \mathbf{k}\right)  }{\left(  \epsilon_{\eta\mathbf{k}}%
-\epsilon_{\eta^{\prime\prime}\mathbf{k}}\right)  ^{2}},
\end{equation}
and
\begin{align}
\left(  \mathbf{A}_{l}^{\text{sj}}\right)  _{\beta}  &  =-2\pi\sum
_{\eta^{\prime}\mathbf{k}^{\prime}}W_{\mathbf{kk}^{\prime}}\delta\left(
\epsilon_{\eta\mathbf{k}}-\epsilon_{\eta^{\prime}\mathbf{k}^{\prime}}\right)
\nonumber\\
&  \times\operatorname{Im}\left[  \sum_{\eta^{\prime\prime}\neq\eta^{\prime}%
}\frac{\langle u_{\eta\mathbf{k}}\mathbf{|}u_{\eta^{\prime}\mathbf{k}^{\prime
}}\rangle\langle u_{\eta^{\prime\prime}\mathbf{k}^{\prime}}\mathbf{|}%
u_{\eta\mathbf{k}}\rangle A_{\beta}^{\eta^{\prime}\eta^{\prime\prime}}\left(
\mathbf{k}^{\prime}\right)  }{\epsilon_{\eta^{\prime}\mathbf{k}^{\prime}%
}-\epsilon_{\eta^{\prime\prime}\mathbf{k}^{\prime}}}\right. \nonumber\\
&  \left.  -\sum_{\eta^{\prime\prime}\neq\eta}\frac{\langle u_{\eta
^{\prime\prime}\mathbf{k}}\mathbf{|}u_{\eta^{\prime}\mathbf{k}^{\prime}%
}\rangle\langle u_{\eta^{\prime}\mathbf{k}^{\prime}}\mathbf{|}u_{\eta
\mathbf{k}}\rangle A_{\beta}^{\eta\eta^{\prime\prime}}\left(  \mathbf{k}%
\right)  }{\epsilon_{\eta\mathbf{k}}-\epsilon_{\eta^{\prime\prime}\mathbf{k}}%
}\right]  . \label{sj-velocity}%
\end{align}
originates from the scattering-induced dressing.
Equation~(\ref{sj-velocity}) is diagrammatically represented in
Fig.~\ref{fig:sj-vertex}. The summation over repeated spatial indices $\alpha
$, $\beta$ is implied hereafter. Here $A_{\beta}^{\eta^{\prime\prime}\eta
}\left(  \mathbf{k}\right)  \equiv\langle u_{\eta^{\prime\prime}\mathbf{k}%
}\mathbf{|}A_{\beta}\mathbf{|}u_{\eta\mathbf{k}}\rangle$ with $\mathbf{|}%
u_{\eta\mathbf{k}}\rangle$ the periodic part of the Bloch state. For
impurities $W_{\mathbf{kk}^{\prime}}=n_{i}\left\vert V_{\mathbf{kk}^{\prime}%
}^{o}\right\vert ^{2}$, with $n_{i}$ the impurity density and $V_{\mathbf{kk}%
^{\prime}}^{o}$ the plane-wave part of the matrix element of the impurity
potential. For electron-phonon scattering%
\begin{equation}
W_{\mathbf{kk}^{\prime}}=\frac{2N_{\mathbf{q}}}{\text{V}}\left\vert
U_{\mathbf{kk}^{\prime}}^{o}\right\vert ^{2},
\end{equation}
where $U_{\mathbf{k}^{\prime}\mathbf{k}}^{o}$ is the plane-wave part of the
electron-phonon matrix element, $N_{\mathbf{q}}$ is the Bose occupation
function of phonons ($\mathbf{q}$ is the wave-vector of phonons with energy
$\hbar\omega_{q}$), V is the volume (area in two-dimension) of the system, and
the factor $2$ accounts for the absorption and emission of phonons.

When calculating the electric current $\mathbf{A}=e\mathbf{v}$, $\Omega
_{\alpha\beta}^{A}$ and $\mathbf{v}_{l}^{\text{sj}}$ are the Berry curvature
and \textquotedblleft side-jump velocity\textquotedblright%
\ \cite{Sinitsyn2007,Sinitsyn2006,Xiao2017SOT-SBE}, respectively. When
calculating the spin current $\mathbf{A}=\mathbf{j}$, $\Omega_{\alpha\beta
}^{A}$ is the so-called spin Berry curvature \cite{Yan2016}, whereas
$\mathbf{A}_{l}^{\text{sj}}$ provides the spin-current counterpart of the
side-jump velocity \cite{Xiao2017SHE,Xiao2017SOT-SBE}. 

The occupation function of the carrier states is decomposed, around the Fermi
distribution $f_{l}^{0}$, into $f_{l}=f_{l}^{0}+g_{l}^{2s}+g_{l}^{a}$.
Its out-of-equilibrium part satisfies the linearized steady-state Boltzmann
equations $e\mathbf{E}\cdot\mathbf{v}_{l}^{0}\partial f_{l}^{0}/\partial\epsilon_{l}=-\sum
_{l^{\prime}}w_{ll^{\prime}}^{2s}\left(  g_{l}^{2s}-g_{l^{\prime}}%
^{2s}\right)  $ and
\begin{equation}
e\mathbf{E}\cdot\mathbf{v}_{l}^{\text{sj}}\partial f_{l}^{0}/\partial
\epsilon_{l}=\sum_{l^{\prime}}w_{ll^{\prime}}^{2s}\left(  g_{l}^{a}%
-g_{l^{\prime}}^{a}\right)  \label{SBE-a}%
\end{equation}
in the presence of weak scalar disorder \cite{Sinitsyn2006}. $w_{ll^{\prime}%
}^{2s}=\frac{2\pi}{\hbar}$ $W_{\mathbf{kk}^{\prime}}\left\vert \langle
u_{l}|u_{l^{\prime}}\rangle\right\vert ^{2}\delta\left(  \epsilon_{l}%
-\epsilon_{l^{\prime}}\right)  $ is the lowest-Born-order scattering rate,
$\mathbf{v}_{l}^{0}$ is the usual band velocity.

Collecting the above ingredients, the spin Hall current is \cite{note-insk,Supp}
$j_{\text{SH}}=j_{\text{SH}}^{\text{bc}}+j_{\text{SH}}^{\text{sj}%
}+j_{\text{SH}}^{\text{ad}}$. The first two terms
\begin{equation}
j_{\text{SH}}^{\text{bc}}=\sum_{l}\left(  j_{l}^{\text{bc}}\right)  f_{l}%
^{0}\text{, and\thinspace\ }j_{\text{SH}}^{\text{sj}}=\sum_{l}\left(
j_{l}^{\text{sj}}\right)  g_{l}^{2s} \label{SBE response}%
\end{equation}
arise from off-shell-states induced corrections to the semiclassical value of
$j_{l}^{0}$. Whereas $j_{\text{SH}}^{\text{ad}}=\sum_{l}j_{l}^{0}g_{l}^{a}$
incorporates the nonequilibrium occupation function modified by off-shell
states, since $g_{l}^{a}$ appears as a response to the generation term
proportional to the \textquotedblleft side-jump velocity\textquotedblright%
\ $\mathbf{v}_{l}^{\text{sj}}$ \cite{Sinitsyn2007}. In calculating the
anomalous Hall (AH) current \cite{Sinitsyn2008} $\mathbf{A}=e\mathbf{v}$,
$j_{\text{AH}}^{\text{sj}}$ and $j_{\text{AH}}^{\text{ad}}$ are related to the
transverse (side-way) and longitudinal components of $\mathbf{v}%
_{l}^{\text{sj}}$, respectively \cite{Sinitsyn2007}. Thereby their sum is also
often referred to as the side-jump contribution in the literature on the
anomalous Hall effect \cite{Sinitsyn2006,Sinitsyn2007,Sinitsyn2008}. Given
this convention, we also include the $j_{\text{SH}}^{\text{ad}}$ contribution,
although $j_{\text{AH}}^{\text{ad}}$ has nothing to do with the original
concept of side-jump, and the microscopic theory \cite{KL1957,Xiao2019NLHE}
indeed shows that $g_{l}^{a}$ is not related to the interband elements of the
out-of-equilibrium density matrix. In fact, in two-dimensional nonmagnetic
models for the spin Hall effect, such as the two-dimensional electronic
systems with Rashba, cubic Rashba and Dresselhaus spin-orbit couplings
\cite{Rashba,Lei2005,Loss2005,Zhang2005,Moriya2014,Culcer2018}, the spin
current operator (for out-of-plane spin component) has only interband matrix
elements, i.e., $j_{l}^{0}=0$, thus $j_{\text{SH}}^{\text{ad}}$\ does not
appear at all \cite{note-insk,Supp}.

\emph{{\color{blue} Phonon-induced T-dependence of spin Hall conductivity.}%
}---In order to show the $T$-dependence of the phonon-induced side-jump spin
Hall conductivity, we prove that its values in the low-$T$ and high-$T$ limits
can be different.

In the low-$T$ limit, $W_{\mathbf{kk}^{\prime}}$\ for phonon scattering is
highly peaked at vanishing scattering angle, and the on-shell scattering can
only be the intraband transition, hence in Eq. (\ref{sj-velocity})
$\eta^{\prime}=\eta$, and $\mathbf{k}^{\prime}$ is very close to $\mathbf{k}$.
We then expand the integrand up to the first order of $\left(  \mathbf{k}%
^{\prime}-\mathbf{k}\right)  $, getting $\left(  \mathbf{A}_{l}^{\text{sj}%
}\right)  _{\beta}=\sum_{\mathbf{k}^{\prime}}\tilde{w}_{l^{\prime}l}%
^{2s}\Omega_{\alpha\beta}^{A}\left(  \eta\mathbf{k}\right)  \left(
\mathbf{k}^{\prime}-\mathbf{k}\right)  _{\alpha}$, with $\tilde{w}_{l^{\prime
}l}^{2s}=\frac{2\pi}{\hbar}W_{\mathbf{kk}^{\prime}}\delta\left(
\epsilon_{\eta\mathbf{k}}-\epsilon_{\eta\mathbf{k}^{\prime}}\right)  $ the
scattering rate in the small-scattering-angle limit. Concurrently,
$\mathbf{v}_{l}^{\text{sj}}=\sum_{\mathbf{k}^{\prime}}\tilde{w}_{l^{\prime}%
l}^{2s}\mathbf{\Omega}\left(  l\right)  $ $\times\left(  \mathbf{k}^{\prime
}-\mathbf{k}\right)  $ ($\mathbf{\Omega}$ is the vector form of the ordinary Berry
curvature) yields $g_{l}^{a}=e\mathbf{E}\cdot\left[  \mathbf{k\times\Omega
}\left(  l\right)  \right]  \partial f_{l}^{0}/\partial\epsilon_{l}$. Thus one
has \cite{Supp}
\begin{equation}
\left(  j_{\text{SH}}^{\text{sj}}\right)  _{\beta}=e\sum_{l}\mathbf{k}%
_{\alpha}\Omega_{\alpha\beta}^{j}\left(  l\right)  \mathbf{E}\cdot
\mathbf{v}_{l}^{0}\partial f_{l}^{0}/\partial\epsilon_{l},
\label{sj-SHC-low-T}%
\end{equation}
and
\begin{equation}
\left(  j_{\text{SH}}^{\text{ad}}\right)  _{\beta}=e\sum_{l}\left(
j_{l}^{0}\right)  _{\beta}\mathbf{E}\cdot\left[  \mathbf{k\times\Omega}\left(
l\right)  \right]  \partial f_{l}^{0}/\partial\epsilon_{l}.
\end{equation}
Thus $\sigma_{\text{SH}}=\left(  j_{\text{SH}}\right)  _{x}/E_{y}$ is a
$T$-independent constant in the low-$T$ limit. It is clear that this constant equals that
contributed by scalar-impurities in the long-range limit, whose
$W_{\mathbf{kk}^{\prime}}$ is also highly concentrated around vanishing
scattering angle.

In the high-$T$ limit, the phonon energy is much smaller than $k_{B}T$
indicating $W_{\mathbf{kk}^{\prime}}=2k_{B}T$V$^{-1}\left\vert U_{\mathbf{k}%
^{\prime}\mathbf{k}}^{o}\right\vert ^{2}/\hbar\omega_{q}$, then we have
$g_{l}^{2s}\sim T^{-1}$, $j_{l}^{\text{sj}}\sim T$, and $g_{l}^{a}\sim T^{0}$,
and consequently,%
\begin{equation}
\rho\sim T,\text{ \ }\sigma_{\text{SH}}\sim T^{0}.
\end{equation}
Accordingly, in practice the high-$T$ limit is identified as the equipartition
regime with linear-in-$T$ resistivity \cite{Ziman1972}. This regime is usually
marked qualitatively by $T>T_{D}$ in textbooks, with $T_{D}$ the Debye
temperature, but can extend practically to about $T>T_{D}/3$ in Pt, Cu and Au
\cite{Ziman1960,Liu2015}, and to about $T>T_{D}/5$ in Al \cite{Ziman1960}.
Besides, it is apparent that the $T$-independent $\sigma_{\text{SH}}$ in the
equipartition regime can be different from that in the low-$T$ limit.

To acquire a more transparent picture, we assume any large-angle
electron-phonon scattering can occur via normal processes, and take the
approximation of the deformation-potential electron-acoustic phonon coupling,
for which a electron-phonon coupling constant can be introduced as
\cite{Abrikosov,Rammer1986} $\lambda^{2}=2$V$^{-1}\left\vert U_{\mathbf{k}%
^{\prime}\mathbf{k}}^{o}\right\vert ^{2}/\hbar\omega_{q}$, hence arriving at
$W_{\mathbf{kk}^{\prime}}=\lambda^{2}k_{B}T$ in the high-$T$ regime. This
$W_{\mathbf{kk}^{\prime}}$ is uniformly distributed on the Fermi surface, just
similar to $W_{\mathbf{kk}^{\prime}}=n_{i}V_{i}^{2}$ contributed by randomly
distributed zero-range scalar impurities, with $V_{i}$ the strength.
Therefore, we infer that the $\sigma_{\text{SH}}$ due to phonons in the
high-$T$ equipartition regime takes the same value as that due to zero-range scalar
impurities. This speculation can be verified by noticing that $g_{l}%
^{2s,ep}\lambda^{2}k_{B}T=g_{l}^{2s,ei}n_{i}V_{i}^{2}$, $\left(
\mathbf{A}_{l}^{\text{sj}}\right)  ^{ep}/\lambda^{2}k_{B}T=\left(
\mathbf{A}_{l}^{\text{sj}}\right)  ^{ei}/n_{i}V_{i}^{2}$ and $g_{l}%
^{a,ep}=g_{l}^{a,ei}$, where the superscripts \textquotedblleft$ep$%
\textquotedblright\ and \textquotedblleft$ei$\textquotedblright\ mean the
contributions due to electron-phonon scattering and zero-range scalar
impurities, respectively.\

According to the above results, $\sigma_{\text{SH}}$ induced by
electron-acoustic phonon scattering is $T$-dependent provided that the
$\sigma_{\text{SH}}$ induced by scalar impurity scattering in the long-range
and zero-range limits are different. This unexpected relation in turn provides
a qualitative picture for comprehending the $T$-dependence of phonon-induced
side-jump, by analogy with the recently revealed sensitivity of the side-jump
conductivity to the scattering range of impurities \cite{Ado2017}. The
accessible phase-space of the electron-phonon scattering changes with
temperature, thus implies a $T$-dependent average momentum transfer, i.e.,
effective range, of this scattering.

Note that this mechanism differs from that for the $T$-dependent $\rho$. To
directly see this point, one need just consider the fact that in the
equipartition regime $\sigma_{\text{SH}}\sim T^{0}$ although $\rho\sim T$ is
still $T$-dependent.

The above revealed relation facilitates judging whether a model system has a
$T$-dependent phonon-induced side-jump conductivity based on the
familiar knowledge about the impurity-induced side-jump.
There are models which possess different side-jump conductivities induced by
scalar impurity-scattering in the long-range and
zero-range limits, such as the Luttinger model describing $p$-type
semiconductor \cite{Murakami2004} and the $k$-cubic Rashba model for the
two-dimensional heavy-hole gas in confined quantum wells \cite{Lei2005}.
In these systems the phonon-induced side-jump conductivities are thus
$T$-dependent.

In the $k$-cubic Rashba model \cite{Moriya2014,Culcer2018}, the Hamiltonian reads
\begin{equation}
\hat{H}=\frac{\hbar^{2}\mathbf{k}^{2}}{2m}+i\frac{\alpha_{R}}{2}\left(
\hat{\sigma}_{+}k_{-}^{3}-\hat{\sigma}_{-}k_{+}^{3}\right)  ,
\end{equation}
where $\mathbf{k}=k\left(  \cos\phi,\sin\phi\right)  $ is the two-dimensional
wave-vector, $k_{\pm}=k_{x}\pm ik_{y}$, $\hat{\sigma}$'s are Pauli matrices
with$\ \hat{\sigma}_{\pm}=\hat{\sigma}_{x}\pm i\hat{\sigma}_{y}$, $\alpha_{R}$
is the spin-orbit coupling coefficient that can be tuned to very large values
by the gate voltage \cite{Culcer2018}. The spin current operator
\cite{Loss2005,Zhang2005} $\hat{\jmath}_{x}=\frac{3\hbar}{2}\frac{1}%
{2}\left\{  \hat{\sigma}_{z},\hat{v}_{x}\right\}  $ has only interband
components, hence $j_{\text{SH}}=j_{\text{SH}}^{\text{bc}}+j_{\text{SH}%
}^{\text{sj}}$. Letting $\eta=\pm$ labels the two Rashba bands, the spin Berry
curvature is $\Omega_{yx}^{j}\left(  \eta\mathbf{k}\right)  =-\eta\frac
{9\hbar^{3}\sin^{2}\phi}{4m\alpha_{R}k^{3}}$, thus $\sigma_{\text{SH}%
}^{\text{bc}}=\left(  j_{\text{SH}}^{\text{bc}}\right)  _{x}/E_{y}%
=\frac{9e\hbar^{2}}{16\pi m\alpha_{R}}\sum_{\eta}\eta k_{\eta}^{-1}$
\cite{Lei2005,Loss2005}, with $k_{\eta}$ the Fermi wave-number of band $\eta$.
The side-jump spin Hall conductivity due to electron-phonon scattering
in the low-$T$ limit is given by Eq. (\ref{sj-SHC-low-T}) as
\begin{equation}
\sigma_{\text{SH}}^{\text{sj}}=\left(  j_{\text{SH}}^{\text{sj}}\right)
_{x}/E_{y}=\frac{1}{4}\sigma_{\text{SH}}^{\text{bc}}.
\end{equation}
When $m\alpha_{R}/\hbar^{2}\ll1/\sqrt{\pi n}$ \cite{Loss2005}, one has
$\sigma_{\text{SH}}^{\text{sj}}=9e/32\pi$. In the high-$T$ regime,
we have $j_{l}^{\text{sj}}=0$ \cite{Supp}, leading to
\begin{equation}
\sigma_{\text{SH}}^{\text{sj}}=0.
\end{equation}
Since the side-jump conductivities in the high-$T$ and low-$T$ limits are
different, there must be a crossover in the intermediate regime resulting in
the $T$-dependent behavior.

\begin{figure}[tbh]
\includegraphics[width=0.7\columnwidth]{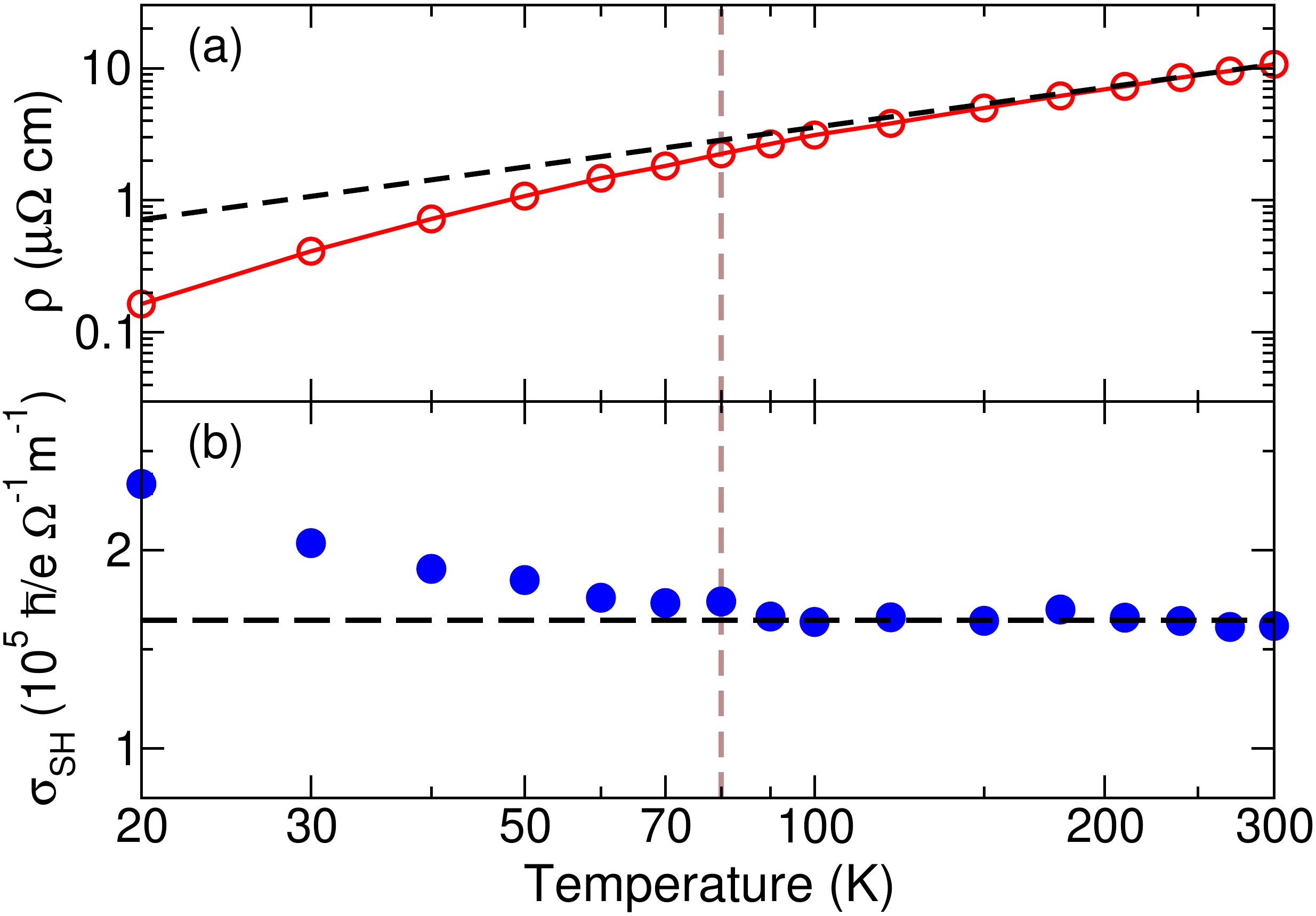}\caption{Calculated
longitudinal resistivity $\rho$ (a) and spin Hall conductivity $\sigma
_{\text{SH}}$ (b) of pure Pt as a function of temperature. The black dashed
line in panel (a) illustrates the linear dependence.}%
\label{fig:Pt}%
\end{figure}

\emph{{\color{blue} Temperature--dependent spin Hall conductivity in pure
Platinum.}}---To show the applicability of our theoretical ideas in real
materials, we perform a first-principles calculation to the spin Hall
conductivity of pure Pt in the range 20 -- 300 K. The minimal interband
splitting around the Fermi level of Pt is much larger than 300 K
\cite{Tanaka2008}, thus the spin Berry-curvature contribution should be
$T$-independent up to 300 K. The temperature is modeled by populating the
calculated phonon spectra of Pt into a large supercell with its length $L$
along fcc [111] and $5\times5$ unit cells in the lateral dimensions
\cite{Liu2015,Starikov2018}. Then the transport calculation is carried out
using the above disordered (finite-temperature) supercell sandwiched by two
perfectly crystalline (zero-temperature) Pt electrodes. The scattering matrix
is obtained using the so-called \textquotedblleft wave function
matching\textquotedblright\ technique within the Landauer-B{\"{u}}ttiker
formalism \cite{Starikov2018}. The calculated total resistance of the
scattering geometry is found to be linearly dependent on $L$ following the
Ohm's law. By varying $L$ in the range of 5 -- 60 nm, we extract the
resistivity at every temperature using a linear least squares fitting for the
calculated resistances. For each $L$, at least 10 random configurations have
been considered to ensure both average value and standard deviation well
converged with respect to the number of configurations. The calculated
resistivity $\rho$ is plotted in Fig.~\ref{fig:Pt}(a) as a function of
temperature. The spin-Hall angle $\Theta_{\mathrm{SH}}$ is computed by
examining the ratio of transverse spin current density and longitudinal charge
current density~\cite{Xia2016}. At every temperature, we use more than 20
random configurations, each of which contains 60 nm-long disordered Pt. Then
the spin Hall conductivity is obtained as $\sigma_{\mathrm{SH}}=(\hbar
/e)\Theta_{\mathrm{SH}}/\rho$, shown in Fig.~\ref{fig:Pt}(b).

For $T\gtrsim80$~K, a linear-in-$T$ $\rho$ is obtained, and $\sigma
_{\mathrm{SH}}$ is approximately a constant of $1.6\times10^{5}~\hbar
/e~(\Omega\,\mathrm{m})^{-1}$ \cite{Xia2016}. Below the equipartition regime,
$\rho$ deviates from the linear $T$-dependence [illustrated by the black
dashed line in Fig.~\ref{fig:Pt}(a)], concurrently the calculated
$\sigma_{\mathrm{SH}}$ increases with decreasing temperature. At $T=20$~K,
$\sigma_{\mathrm{SH}}$ reaches $2.3\times10^{5}~\hbar/e~(\Omega\,\mathrm{m}%
)^{-1}$. Compared to the value at 80 K, the $T$-variation of $\sigma
_{\mathrm{SH}}$ is as large as 50\%. This $T$-dependence just begins when the
temperature drops below the equipartition regime, in agreement with our
theoretical prediction.

Finally, we discuss the possibility of observing the predicted effect in
experiments. In high-purity metals, the electron-electron scattering dominates
over the electron-phonon scattering in determining transport behaviors at very
low temperature. To observe our prediction, lower characteristic temperature
$T_{t}$ marking the crossover from the electron-electron dominated regime to
the electron-phonon dominated one is required, such that the intermediate
range from $T_{t}$ to the high-$T$ equipartition regime is wide enough. In
experimentally accessible high-purity Pt samples with residual resistivity as
small as $10^{-3}$ -- $10^{-2}$ $\mu\Omega$ cm \cite{Bass1969,White1958},
$T_{t}$ can be as low as 10K, and at $T=20$ K the phonon-induced $\rho$ is
nearly one order of magnitude larger than that contributed by the
electron-electron scattering and the residual resistivity \cite{Bass1969}.
Because in high-purity Pt, the $T$-linear scaling of $\rho$ emerges at
$T\gtrsim80$ K \cite{Liu2015}, the suitable range for observing the
first-principles predicted $T$-dependence of $\sigma_{\mathrm{SH}}$
[Fig.~\ref{fig:Pt}(b)] is $20$ K $\lesssim T\lesssim80$ K. Very recently
experimentalists have been developing new techniques, with which the spin
current is generated and detected in a single transition-metal sample, thus
avoiding all the complications associated with the interfaces and shunting
effect \cite{Stamm2017,Jin2019}. The predicted effect is expected to be
observed as the quality of Pt samples in such measurements is improved.

\begin{acknowledgments}
We thank Ming Xie, Tianlei Chai and Haodi Liu for helpful discussions.
Q.N. is supported by DOE (DE-FG03-02ER45958, Division of Materials Science and Engineering) on the transport formulation in this work. C.X. is supported by NSF (EFMA-1641101) and Welch Foundation (F-1255). S.A.Y. is supported by Singapore Ministry of Education AcRF Tier 2 (MOE2017-T2-2-108). Y.L. and Z.Y are supported by the National Natural Science Foundation of China (Grants No. 61604013, No. 61774018, and No. 11734004), the Recruitment Program of Global Youth Experts, and the Fundamental Research Funds for the Central Universities (Grants No. 2016NT10 and No. 2018EYT03).
\end{acknowledgments}


\begin{thebibliography}{99}                                                                                               %


\bibitem {Sinova2015}J. Sinova, S. O. Valenzuela, J. Wunderlich, C. H. Back,
and T. Jungwirth, Rev. Mod. Phys. \textbf{87}, 1213 (2015).

\bibitem {Hoffmann2013}A. Hoffmann, IEEE Trans. Magn. \textbf{49}, 5172 (2013).

\bibitem {Seki2008}T. Seki, Y. Hasegawa, S. Mitani, S. Takahashi, H. Imamura,
S. Maekawa, J. Nitta, and K. Takanashi, Nat. Mater. \textbf{7}, 125 (2008).

\bibitem {Hoffmann2010}O. Mosendz, J. E. Pearson, F. Y. Fradin, G. E.W. Bauer,
S. D. Bader, and A. Hoffmann, Phys. Rev. Lett. \textbf{104}, 046601 (2010).

\bibitem {Liu2012}L. Liu, C.-F. Pai, Y. Li, H.W. Tseng, D. C. Ralph, and R. A.
Buhrman, Science \textbf{336}, 555 (2012).

\bibitem {Sagasta2016}E. Sagasta, Y. Omori, M. Isasa, M. Gradhand, L. E.
Hueso, Y. Niimi, Y. C. Otani, and F. Casanova, Phys. Rev. B \textbf{94},
060412(R) (2016).

\bibitem {Stamm2017}C. Stamm, C. Murer, M. Berritta, J. Feng, M. Gabureac, P.
M. Oppeneer, and P. Gambardella, Phys. Rev. Lett. \textbf{119}, 087203 (2017).

\bibitem {Jin2019}C. Chen, D. Tian, H. Zhou, D. Hou, and X. Jin, Phys. Rev.
Lett. \textbf{122}, 016804 (2019).

\bibitem {Tanaka2008}T. Tanaka, H. Kontani, M. Naito, T. Naito, D. S.
Hirashima, K. Yamada, and J. Inoue, Phys. Rev. B \textbf{77}, 165117 (2008).

\bibitem {Xia2016}L. Wang, R. J. H. Wesselink, Y. Liu, Z. Yuan, K. Xia, and P.
J. Kelly, Phys. Rev. Lett. \textbf{116}, 196602 (2016).

\bibitem {Yan2016}Y. Sun, Y. Zhang, C. Felser, and B. Yan, Phys. Rev. Lett.
\textbf{117}, 146403 (2016).

\bibitem {Berger1970}L. Berger, Phys. Rev. B \textbf{2}, 4559 (1970).

\bibitem {Bruno2001}A. Crepieux and P. Bruno, Phys. Rev. B \textbf{64}, 014416 (2001).

\bibitem {Ziman1972}J. M. Ziman, {\itshape}Principles of the Theory of Solids
(Cambridge University Press, Cambridge, 1972).

\bibitem {Liu2015}Y. Liu, Z. Yuan, R. J. H. Wesselink, A. A. Starikov, M. van
Schilfgaarde, and P. J. Kelly, Phys. Rev. B \textbf{91}, 220405(R) (2015).

\bibitem {Lyo1972}S. K. Lyo and T. Holstein, Phys. Rev. Lett. \textbf{29}, 423 (1972).

\bibitem {Engel2005}H.-A. Engel, B. I. Halperin, and E. I. Rashba, Phys. Rev.
Lett. \textbf{95}, 166605 (2005).

\bibitem {Sarma2006}W. K. Tse and S. Das Sarma, Phys. Rev. Lett. \textbf{96},
056601 (2006).

\bibitem {Vignale2006}E. M. Hankiewicz and G. Vignale, Phys. Rev. B
\textbf{73}, 115339 (2006).

\bibitem {KL1957}W. Kohn and J. M. Luttinger, Phys. Rev. \textbf{108}, 590
(1957); J. M. Luttinger, Phys. Rev. \textbf{112}, 739 (1958).

\bibitem {Sinitsyn2008}N. A. Sinitsyn, J. Phys.: Condens. Matter \textbf{20},
023201 (2008).

\bibitem {Sinitsyn2006}N. A. Sinitsyn, Q. Niu, and A. H. MacDonald, Phys. Rev.
B 73, 075318 (2006).

\bibitem {Xiao2019NLHE}C. Xiao, Z. Z. Du, and Q. Niu, arXiv: 1907.00577

\bibitem {Xiao2017SOT-SBE}C. Xiao and Q. Niu, Phys. Rev. B \textbf{96}, 045428 (2017).

\bibitem {Sinitsyn2007}N. A. Sinitsyn, A. H. MacDonald, T. Jungwirth, V. K.
Dugaev, and J. Sinova, Phys. Rev. B \textbf{75}, 045315 (2007).

\bibitem {Xiao2017SHE}C. Xiao, Front. Phys. \textbf{13}, 137202 (2018).

\bibitem {note-insk}For our purpose of showing the $T$-dependence of the
spin Hall conductivity induced by electron-phonon scattering, the results
for $j_{\text{SH}}^{\text{sj}}$ and $j_{\text{SH}}^{\text{ad}}$ are sufficient.
For completeness, we present in Ref. \cite{Supp} the other spin Hall contribution
of order $\left(  W_{\mathbf{kk}^{\prime}}\right)  ^{0}$ from the
antisymmetric part of the third-Born-order scattering rate. This contribution
also vanishes when $j_{l}^{0}=0$.

\bibitem {Supp}See Supplemental Material for the transport formalism for the
spin Hall conductivities of order $\left(  W_{\mathbf{kk}^{\prime}}\right)^{0}$
and calculation details of the spin Hall conductivity in the $k$-cubic Rashba model,
as well as discussions on the low-$T$ limit in the presence of both electron-phonon
and electron-impurity scattering.

\bibitem {Rashba}O. V. Dimitrova, Phys. Rev. B \textbf{71}, 245327 (2005).

\bibitem {Lei2005}S. Y. Liu and X. L. Lei, Phys. Rev. B \textbf{72}, 155314 (2005).

\bibitem {Loss2005}J. Schliemann and D. Loss, Phys. Rev. B \textbf{71}, 085308 (2005).

\bibitem {Zhang2005}B. A. Bernevig and S. C. Zhang, Phys. Rev. Lett.
\textbf{95}, 016801 (2005).

\bibitem {Moriya2014}R. Moriya, K. Sawano, Y. Hoshi, S. Masubuchi, Y. Shiraki,
A. Wild, C. Neumann, G. Abstreiter, D. Bougeard, T. Koga, and T. Machida,
Phys. Rev. Lett. \textbf{113}, 086601 (2014).

\bibitem {Culcer2018}H. Liu, E. Marcellina, A. R. Hamilton, and D. Culcer,
Phys. Rev. Lett. \textbf{121}, 087701 (2018).

\bibitem {Ziman1960}J. M. Ziman, {\itshape}Electrons and Phonons (Clarendon,
Oxford, 1960).

\bibitem {Abrikosov}A. A. Abrikosov, L. P. Gor'kov, and I. E. Dzyaloshinskii,
{\itshape}Quantum Field Theoretical Methods in Statistical Physics (Pergamon
Press, New York, 1965).

\bibitem {Rammer1986}J. Rammer and H. Smith, Rev. Mod. Phys. \textbf{58}, 323 (1986).

\bibitem {Ado2017}I. A. Ado, I. A. Dmitriev, P. M. Ostrovsky, and M. Titov,
Phys. Rev. B \textbf{96}, 235148 (2017).

\bibitem {Murakami2004}S. Murakami, Phys. Rev. B \textbf{69}, 241202(R) (2004).

\bibitem {Starikov2018}Anton A. Starikov, Yi Liu, Zhe Yuan and Paul J. Kelly,
Phys. Rev. B \textbf{97}, 214415 (2018).

\bibitem {White1958}G. K. White and S. B. Woods, Phil. Trans. Roy. Soc.
(London) \textbf{A251}, 273 (1958).

\bibitem {Bass1969}R. H. Freeman, F. J. Blatt, and J. Bass, Phys. Kondens.
Materie \textbf{9}, 271 (1969).
\end{thebibliography}
\end{document}